\documentclass{article}
\PassOptionsToPackage{hyphens}{url}
\usepackage[sglblindworkshop,final]{neurips_2026}
\workshoptitle{Machine Learning for Systems}

\makeatletter
\renewcommand{\@noticestring}{%
  The 10th Machine Learning for Systems Workshop at NeurIPS 2026, Sydney, Australia%
}
\makeatother

\IfFileExists{phvr8t.tfm}{}{}
\IfFileExists{ectt1000.tfm}{}{}

\usepackage[utf8]{inputenc}
\usepackage[T1]{fontenc}
\usepackage{hyperref}
\usepackage{url}
\usepackage{booktabs}
\usepackage{amsfonts}
\usepackage{amsmath}
\usepackage{microtype}
\usepackage{xcolor}
\usepackage{comment}

\IfFileExists{siunitx.sty}{\usepackage{siunitx}}{%
  \providecommand{\mega}{M}%
  \providecommand{\kilo}{k}%
  \providecommand{\hertz}{Hz}%
  \providecommand{\SI}[2]{##1\,##2}%   parameters doubled: \IfFileExists expands its argument
}
\usepackage{graphicx}
\title{Workload Identification with Physical Side Channels for AI Governance}

\author{%
  Simone Gargiulo \\
  Pivotal Research \\
  \texttt{simonegargiulo2@gmail.com}\\
  \And
  Gabriel Kulp \\
  Intelligence Security Laboratories \\
 \texttt{gabriel@intseclab.org}\\
}
\begin{document}
\maketitle

\begin{abstract}
AI compute verification is one of the first tangible and tractable points for international policy aimed at AI governance. Determining whether frontier labs, or any operator, comply with established agreements requires the regulating authority, at the very least, to discern how their compute is used.
The elementary building block of AI compute is the GPU, and any activity it executes leaves a physical trace.
Here, we show that an external observer can identify the class of the workload running on a single NVIDIA H200 from its power draw. Unlike on-chip NVML telemetry, which can be spoofed or replayed, this physical channel can be observed without the operator's cooperation.
We recorded $930$ five-second traces at \SI{10}{\mega\hertz} with a Rogowski probe, covering seventeen open-weight LLMs and twenty-five non-AI workloads. 
Over this corpus, we separate training from inference and from non-AI computation with an accuracy of $97\%$ and a macro-averaged F1 score of $0.955$, evaluated on models unseen during training. We find that AI workload spectral content predominantly lies below $\sim\SI{20}{\kilo\hertz}$ and training appears to be particularly recognizable through the optimizer update.
The GPU operator is then treated as adversarial and able to reshape the training program. Four evasion strategies are tested with the aim of disguising training as inference, producing an additional 680 adversarial traces.  
A detector hardened against evasion strategies, with the tested strategy held out, catches training $\geq 99\%$ of the time for three of the four strategies. The fourth, diluted low-rank adaptation (LoRA), is detected $48$--$88\%$ of the time with a hardened classifier, rising to $\geq 98\%$ with an additional rescue rule. While these attacks are not intended to provide a comprehensive evaluation against adversarial behaviour, they offer initial insights beyond genuine activities and a dataset for developing and testing stronger evasion mechanisms.

\end{abstract}

\section{Introduction}
\label{sec:intro}

Proposals to regulate large-scale training of AI models rely on a verifier's ability to establish the activity of a cluster, without trusting the operator's self-report~\cite{shavit2023chinchilla,scher2024mechanisms,baker2025verifying,reuel2025open}. Therefore, identifying the class of workload a GPU is running, without direct access to the software stack, is a measurement system problem of increasing importance for international policy.

Several physical parameters are affected by the workload being executed on the GPU. Recently, on-chip NVML telemetry, sampled at \SI{1}{\hertz}, has been shown to reliably detect LLM training when the operator is non-adversarial~\cite{rahman2026detecting}. The picture becomes more complex when the operator actively tries to manipulate the physical signature to disguise training as inference. The classifier proposed in the study has been exposed to several rounds of red-teaming across 20 evasion strategies, including white-box diluted LoRA, which reduces the detection rate to~\SI{42.6}{\percent}. Compared with full fine-tuning, diluted LoRA interleaves small adapter weight updates with inference, hiding one of the training fingerprints within a window mostly occupied by inference, at the cost of reduced training throughput.

Beyond the detection rate itself, the main vulnerability of on-chip telemetry counters is that they are read by software under the operator's control. In fact, they could be reported without being measured, or replayed from an authorized workload while a hidden one is running.
If the operator is capable of this manipulation and willing to accept some performance penalty, no amount of classifier hardening can detect it. Research efforts should therefore focus on tamper-resistant read paths and attested telemetry channels, as suggested by work on hardware-enabled mechanisms~\cite{kulp2024hardware,aarne2024secure,ogara2025hardware,petrie2025flexheg}.

On-chip (NVML) telemetry reports several workload-dependent signals, e.g. power, temperature, clock speed, and PCIe throughput, with updates at tens of Hertz. An external probe, owned by the auditor, can read similar underlying signals, such as current draw, at much higher sampling rates in the \si{\mega\hertz} range, trading the multidimensionality of slow telemetry counters for a single channel with much higher temporal resolution. Because it bypasses the operator's control, this channel can in principle be read without trusting self-reported values and can serve as an additional defence layer or feed a sensor-fusion analysis. Deploying such a probe would require regulation of its own, together with tamper-resistant mechanisms that guarantee the fidelity of its readout. 

Physical side channels have been widely reported in the literature, mostly in the form of attacks. For example, they have been used to obtain sensitive information about the execution of ML models: to recover neural-network architectures and weights~\cite{maia2022can,xiao2026peering,horvath2026kraken}, and to reconstruct user text from mixture-of-experts (MoE) routing~\cite{ding2025moecho,nuriyev2026expert}.
The same physics has also been considered for defensive monitoring~\cite{DARPA_LADS}, for example to detect anomalous program execution on embedded devices through electromagnetic emissions~\cite{sehatbakhsh2019emma}, or to flag cryptojacking through magnetic leakage signals~\cite{xiao2023magtracer}.

Beyond~\cite{rahman2026detecting}, NVML has also been used to profile workload power for whole-facility infrastructure and energy planning~\cite{vercellino2026measurement}. 
In this work, we consider a verifier-owned current probe on the auxiliary power supply of a single GPU, while the operator retains full control of the GPU. With this probe, we record $1610$ traces ($930$ genuine and $680$ adversarial) across seventeen open-weight models and twenty-five non-AI workloads.

\section{Measurement and corpus}
\label{sec:setup}

Workload activity is recorded using a Rogowski current probe~\cite{micsig_rcp} clamped around all positive conductors of the PCIe auxiliary power supply of a single NVIDIA H200 NVL hosted by~\cite{sidechannelcloud}, capturing the magnetic field generated by the aggregate current drawn by the GPU. The probe is AC-coupled, has no DC response and a \SI{34}{\hertz} to \SI{30}{\mega\hertz} bandwidth. Output is digitized using a PicoScope at a nominal sampling rate of \SI{10}{\mega\hertz} in \SI{5}{\second} acquisition windows, while each workload is executed continuously in a loop. A preliminary acquisition at \SI{100}{\mega\hertz} showed that the spectrum was largely confined below \SI{1}{\mega\hertz}, motivating the lower sampling rate used for the main dataset.
The genuine dataset contains $930$ traces covering seventeen models, ranging from $4$B to $21$B parameters and spanning both dense and mixture-of-experts architectures. Each model is recorded during inference and under three training configurations: full fine-tuning, low-rank adaptation (LoRA), and gradient accumulation. Each configuration is acquired over 10 independent runs, each with a fresh model load, for each of the 17 models ($170$ recordings per training configuration), so that measurements are not conditioned by previous GPU execution state and scheduling history. For the gradient accumulation, the number of micro-batches for each model is chosen to fit a complete accumulation cycle within the acquisition window regardless of its starting phase. Where necessary, we reduced the batch size to prevent the accumulated cycle from exceeding memory, as described in Table~\ref{tab:accumsettings}. Every training trace in the corpus contains at least one complete optimizer update in its acquisition window. We also recorded 25 non-AI workloads, including the idle GPU, which serves as a baseline. The adversarial corpus adds $680$ recordings, corresponding to four evasion strategies across the same 17 models, with 10 independent captures per model.  Appendix~\ref{app:Appendix} provides the full list of workloads and configurations. We release all recordings as a dataset~\cite{dataset}.

\section{Results}
\label{sec:results}

As shown similarly in the telemetry study, a two-stage random forest was found to be among the best approaches for distinguishing different workloads. The detector uses $30$ features listed in Table~\ref{tab:features}, predominantly in the frequency domain, that do not depend on the absolute signal amplitude.

The classifier first asks whether the workload is an AI-related activity, and then whether it is training or inference. We report accuracy and macro-averaged F1, where the latter is computed with equal weight over the three classes regardless of class size. Performances are evaluated with five-fold cross-validation grouped by model. This is repeated over four independent shuffles and two seeds to capture variation in both fold composition and training seed. The reported value is the average over all runs. Detector performance is only weakly affected by the acquisition rate, even after a 256-fold decimation of the recorded traces, as shown in Table~\ref{tab:rates}. This is consistent with most of the relevant content being concentrated at low frequencies.

We then consider the threat model where the operator can manipulate workload execution through four evasion strategies that reshape the usual training pattern to avoid detection: (i) chunking the optimizer update, (ii) throttling, (iii) diluting real steps with decoy inference, and (iv) diluting LoRA. We assume that the verifier can observe only the signal leaking from the power supply. Each strategy is run on all seventeen models and carries a performance penalty. We quantify this penalty as the reduction in training throughput, measured in real training tokens per second, relative to a full fine-tune of the same model. Across the four strategies in Table~\ref{tab:penalty}, the penalty ranges from $28.1\%$ to $69.3\%$.

\begin{figure}[!htbp]
\includegraphics[width=\textwidth]{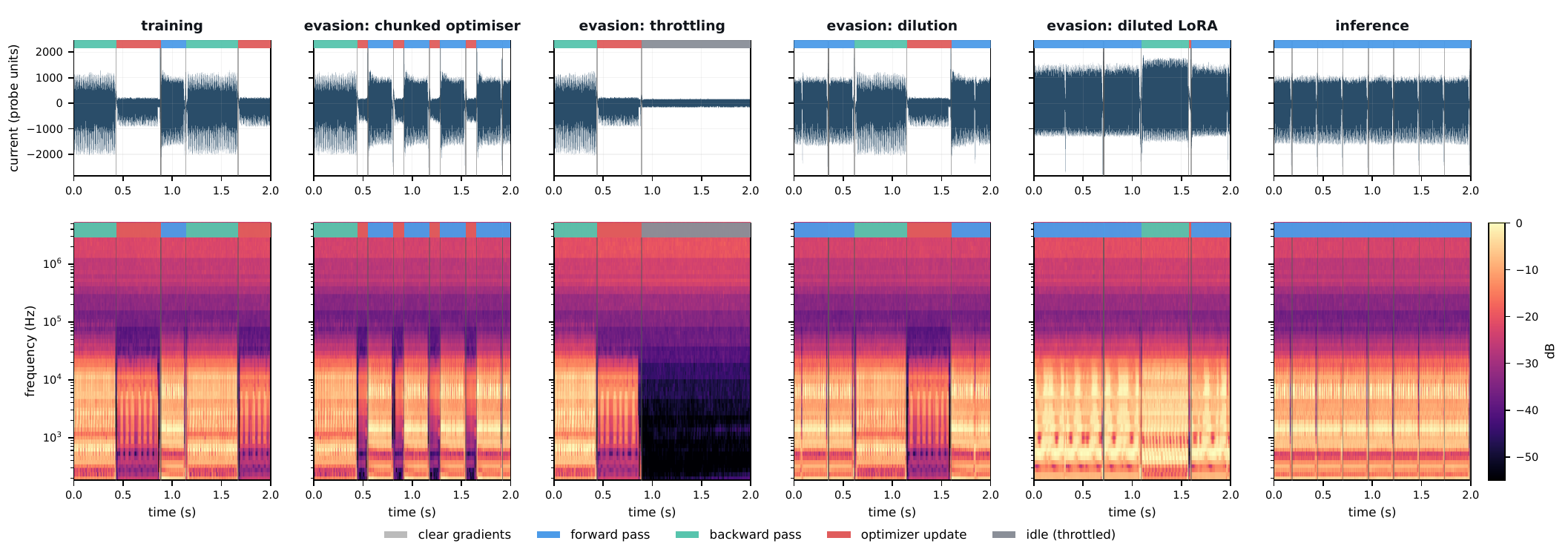}
\caption{Temporal and spectral content of Qwen3-14B traces over an identical \SI{2}{\second} window. From left to right: genuine training, the four evasion strategies, and inference. In the diluted LoRA panel, the optimizer update occupies only a short time window before the model returns to the inference pass. Each time-frequency map is normalized to its own peak within the window.}
\label{fig:evasions}
\end{figure}

Figure~\ref{fig:evasions} shows genuine training and inference recordings for Qwen3-14B, and how each evasion strategy reshapes both the temporal and spectral dynamics of the computation.
A training step is divided into three phases: the forward pass, the backward pass, and the optimizer update. The last two are absent during inference. The optimizer update is what makes training stand out the most. It is expected to be limited by memory bandwidth, which makes it quieter than the other steps and gives it a distinct spectral signature. 

We distinguish three settings for the classifier. The \emph{genuine} detector is trained on the initial 930 traces and has never seen an evasion strategy, providing a baseline for how often training is recognized. The \emph{hardened unseen-strategy} detector is additionally trained on three evasion strategies and tested on the fourth (leave-one-strategy-out). In both settings, all models appear in the training set. The \emph{hardened unseen-model} detector, instead, has seen every strategy but is evaluated on models held out from its training.

%%% TABLE %%%%
\begin{table}[!htbp]
\caption{Training detection rate of the two-stage forest. \emph{Genuine} is trained on the initial $930$ recordings alone. \emph{Unseen strategy} adds the $510$ adversarial recordings from the other three evasion strategies in a leave-one-strategy-out setting. Both detectors see every model. \emph{Unseen model} reverses the split, holding out models rather than evasion strategies. It is trained on all recordings, genuine and adversarial, of the other models, and scored on the traces of the held-out model. The first two columns hold out only the ten recordings of the corresponding label, while the model remains visible; the third holds out all traces associated with a model. Ranges are $95\%$ bootstrap intervals over the seventeen models, averaged over four training seeds.}
\label{tab:detectors}
\centering
\begin{tabular}{lccc}
\toprule
 & & \multicolumn{2}{c}{hardened} \\
\cmidrule(lr){3-4}
Strategy & genuine & \textbf{unseen strategy} & unseen model \\
\midrule
genuine training (baseline) & $94$--$100\%$ & {\boldmath$100\%$} & $100\%$ \\
\midrule
chunked optimizer  & $100\%$      & {\boldmath$100\%$} & $100\%$ \\
throttling         & $74$--$92\%$ & {\boldmath$100\%$} & $100\%$ \\
dilution           & $80$--$100\%$ & {\boldmath$99$--$100\%$} & $100\%$ \\
diluted LoRA       & $30$--$74\%$ & {\boldmath$48$--$88\%$} & $86$--$100\%$ \\
{\quad with rescue rule} & $90$--$100\%$ & {\boldmath{$98$--$100\%$}} & {$100\%$} \\
\bottomrule
\end{tabular}
\end{table}
%%% END OF TABLE %%%%

Hardening improves detection across all strategies, from \SIrange{80}{100}{\percent} to \SI{\geq 99}{\percent} for dilution and from \SIrange{30}{74}{\percent} to \SIrange{48}{88}{\percent} for diluted LoRA. The unseen-model evaluation shows that the main challenge is represented by the novelty of the evasion strategy rather than the variation across models. Values are means over four training seeds.

To improve the detection of diluted LoRA, we introduce a rescue rule targeting the signature left by the optimizer update, which produces a brief, deep power dip with characteristic low-frequency content. Traces labelled as inference by the second random-forest stage can be relabelled as training when the rescue rule detects this frequency-gated dip. The threshold is calibrated so that no genuine inference or non-AI workloads are relabelled as training. This raises diluted LoRA detection to \SIrange{90}{100}{\percent} with the genuine detector and \SIrange{98}{100}{\percent} with the hardened detector. Appendix~\ref{app:rescue} provides the calibration details and an out-of-fold evaluation of the resulting false-alarm rate. 

\section{Discussion}
\label{sec:discussion}

The on-chip study~\cite{rahman2026detecting} reported, under leave-one-strategy-out evaluation, a mean accuracy of $98.2\%$ across twenty strategy families, including classifier-assisted white-box attacks, and $42.6\%$ on its hardest case, that is white-box LoRA with dilution. Our corresponding strategy is diluted LoRA, for which the hardened detector reaches $48$--$88\%$ across 17 models and $\geq 98\%$ with the rescue rule. These results are obtained from a \SI{5}{\second} recording, compared with a \SI{15}{\second}--\SI{60}{\second} window, under the same leave-one-strategy-out protocol. The higher bandwidth of the Rogowski probe enables detection over much shorter windows and appears to make these evasions substantially more difficult.

While telemetry provides privacy-preserving, zero-overhead signals, it cannot be trusted alone against an adversary who controls both the hardware and software stack, since the reported telemetry could be forged or replayed. An auditor-owned external probe can provide an additional layer of security, but requires physical access to the monitored hardware and likely continuous on-site monitoring by the regulatory authority. This additional infrastructural overhead comes with an important advantage: unlike telemetry, evading this physical side channel requires a substantial modification of the workload execution. 

Several limitations of this study motivate future work. First, the corpus was collected on a single NVIDIA H200 NVL, and generalization across devices and hardware architectures remains to be established. Still, the on-chip NVML results are encouraging, suggesting that the underlying fingerprints may be robust enough to support detection across different hardware.
Second, the adversarial corpus should be extended to include more fine-grained manipulations of the physical execution pattern. We explored only a limited set of evasion strategies, among which diluted LoRA was the most difficult to detect. LoRA updates only a small fraction of the model parameters compared with full fine-tuning, leaving a shorter temporal fingerprint of the optimizer update and therefore making detection more difficult. This evasion comes at the cost of reduced optimization flexibility, since the set of parameters that can be updated is constrained. An important direction for future work is therefore to identify evasion strategies that retain the optimization flexibility of full fine-tuning while still reducing detectability.

\label{endofbody}

\section*{Acknowledgements}

The measurements reported here were collected on an NVIDIA H200 NVL provided by~\cite{sidechannelcloud}. The authors would like to thank Carl Heimann of Amodo Design and acknowledge support from Pivotal Research.

\bibliographystyle{plainnat}
\bibliography{references}

@techreport{kulp2024hardware,
  author      = {Gabriel Kulp and Daniel Gonzales and Everett Smith and Lennart Heim and Prateek Puri and Michael J. D. Vermeer and Zev Winkelman},
  title       = {Hardware-Enabled Governance Mechanisms: Developing Technical Solutions to Exempt Items Otherwise Classified Under Export Control Classification Numbers {3A090} and {4A090}},
  type        = {Working Paper},
  number      = {WR-A3056-1},
  institution = {RAND Corporation},
  address     = {Santa Monica, CA},
  month       = jan,
  year        = {2024},
  url         = {https://www.rand.org/pubs/working_papers/WRA3056-1.html}
}

@article{shavit2023chinchilla,
  author  = {Yonadav Shavit},
  title   = {What Does It Take to Catch a {Chinchilla}? {Verifying} Rules on Large-Scale Neural Network Training via Compute Monitoring},
  journal = {arXiv preprint arXiv:2303.11341},
  year    = {2023},
  doi     = {10.48550/arXiv.2303.11341},
  url     = {https://arxiv.org/abs/2303.11341}
}

@techreport{scher2024mechanisms,
  title       = {Mechanisms to Verify International Agreements About {AI} Development},
  author      = {Scher, Aaron and Thiergart, Lisa},
  institution = {Machine Intelligence Research Institute, Technical Governance Team},
  year        = {2024},
  month       = nov,
  eprint      = {2506.15867},
  archivePrefix = {arXiv},
  primaryClass  = {cs.CY},
  url         = {https://techgov.intelligence.org/research/mechanisms-to-verify-international-agreements-about-ai-development}
}

@techreport{baker2025verifying,
  title       = {Verifying International Agreements on {AI}: Six Layers of Verification for Rules on Large-Scale {AI} Development and Deployment},
  author      = {Baker, Mauricio and Kulp, Gabriel and Marks, Oliver and Brundage, Miles and Heim, Lennart},
  institution = {RAND Corporation},
  type        = {Working Paper},
  number      = {WR-A4077-1},
  address     = {Santa Monica, CA},
  year        = {2025},
  doi         = {10.7249/WRA4077-1},
  eprint      = {2507.15916},
  archivePrefix = {arXiv},
  primaryClass  = {cs.CY},
  url         = {https://www.rand.org/pubs/working_papers/WRA4077-1.html}
}

@article{reuel2025open,
  title   = {Open Problems in Technical {AI} Governance},
  author  = {Reuel, Anka and Bucknall, Ben and Casper, Stephen and Fist, Tim and Soder, Lisa and Aarne, Onni and Hammond, Lewis and Ibrahim, Lujain and Chan, Alan and Wills, Peter and Anderljung, Markus and Garfinkel, Ben and Heim, Lennart and Trask, Andrew and Mukobi, Gabriel and Schaeffer, Rylan and Baker, Mauricio and Hooker, Sara and Solaiman, Irene and Luccioni, Alexandra Sasha and Rajkumar, Nitarshan and Mo{\"e}s, Nicolas and Ladish, Jeffrey and Bau, David and Bricman, Paul and Guha, Neel and Newman, Jessica and Bengio, Yoshua and South, Tobin and Pentland, Alex and Koyejo, Sanmi and Kochenderfer, Mykel J. and Trager, Robert},
  journal = {arXiv preprint arXiv:2407.14981},
  year    = {2025},
  eprint  = {2407.14981},
  archivePrefix = {arXiv},
  primaryClass  = {cs.CY},
  doi     = {10.48550/arXiv.2407.14981},
  note    = {v1 July 2024; v2 April 2025. Authors ask to be cited as ``Reuel, Bucknall, et al. (2025)''}
}

@inproceedings{rahman2026detecting,
  title     = {Detecting Hidden {ML} Training With Zero-Overhead Telemetry},
  author    = {Rahman, Robi and Tajdari, Sabiha},
  booktitle = {Proceedings of the Second Workshop on Technical AI Governance Research (TAIGR) at the 43rd International Conference on Machine Learning},
  year      = {2026},
  address   = {Seoul, South Korea},
  eprint    = {2606.19262},
  archivePrefix = {arXiv},
  primaryClass  = {cs.LG},
  url       = {https://arxiv.org/abs/2606.19262}
}

@article{vercellino2026measurement,
  title   = {Measurement of Generative {AI} Workload Power Profiles for Whole-Facility Data Center Infrastructure Planning},
  author  = {Vercellino, Roberto and Willard, Jared and Campos, Gustavo and da Silva Pereira, Weslley and Hull, Olivia and Selensky, Matthew and Mueller, Juliane},
  journal = {arXiv preprint arXiv:2604.07345},
  year    = {2026},
  eprint  = {2604.07345},
  archivePrefix = {arXiv},
  primaryClass  = {eess.SY},
  url     = {https://arxiv.org/abs/2604.07345}
}

@techreport{aarne2024secure,
  title       = {Secure, Governable Chips: Using On-Chip Mechanisms to Manage National Security Risks from {AI} and Advanced Computing},
  author      = {Aarne, Onni and Fist, Tim and Withers, Caleb},
  institution = {Center for a New American Security},
  type        = {Report},
  year        = {2024},
  month       = jan,
  url         = {https://s3.us-east-1.amazonaws.com/files.cnas.org/documents/CNAS-Report-Tech-Secure-Chips-Jan-24-finalb.pdf}
}

@inproceedings{ogara2025hardware,
  title     = {Hardware-Enabled Mechanisms for Verifying Responsible {AI} Development},
  author    = {O'Gara, Aidan and Kulp, Gabriel and Hodgkins, Will and Petrie, James and Immler, Vincent and Aysu, Aydin and Basu, Kanad and Bhasin, Shivam and Picek, Stjepan and Srivastava, Ankur},
  booktitle = {ICML 2025 Workshop on Technical AI Governance (TAIG)},
  year      = {2025},
  address   = {Vancouver, Canada},
  pages     = {1--7},
  note      = {Spotlight talk},
  eprint    = {2505.03742},
  archivePrefix = {arXiv},
  primaryClass  = {cs.CR},
  doi       = {10.48550/arXiv.2505.03742},
  url       = {https://arxiv.org/abs/2505.03742}
}

@techreport{petrie2025flexheg,
  title       = {Flexible Hardware-Enabled Guarantees for {AI} Compute},
  author      = {Petrie, James and Aarne, Onni and Ammann, Nora and Dalrymple, David},
  institution = {Commissioned by ARIA},
  type        = {Report},
  note        = {Part I of a three-part exploration of flexHEGs},
  year        = {2025},
  month       = apr,
  eprint      = {2506.15093},
  archivePrefix = {arXiv},
  primaryClass  = {cs.CR},
  doi         = {10.48550/arXiv.2506.15093},
  url         = {https://arxiv.org/abs/2506.15093}
}

@inproceedings{maia2022can,
  title     = {Can One Hear the Shape of a Neural Network?: Snooping the {GPU} via Magnetic Side Channel},
  author    = {Maia, Henrique Teles and Xiao, Chang and Li, Dingzeyu and Grinspun, Eitan and Zheng, Changxi},
  booktitle = {31st USENIX Security Symposium (USENIX Security 22)},
  year      = {2022},
  address   = {Boston, MA, USA},
  publisher = {USENIX Association},
  eprint    = {2109.07395},
  archivePrefix = {arXiv},
  url       = {https://www.usenix.org/conference/usenixsecurity22/presentation/maia}
}

@inproceedings{xiao2026peering,
  title     = {Peering Inside the Black-Box: Long-Range and Scalable Model Architecture Snooping via {GPU} Electromagnetic Side-Channel},
  author    = {Xiao, Rui and Feng, Sibo and Ramesh, Soundarya and Han, Jun and Han, Jinsong},
  booktitle = {Proceedings of the 2026 Network and Distributed System Security Symposium (NDSS)},
  year      = {2026},
  note      = {Distinguished Paper Award},
  url       = {https://www.ndss-symposium.org/ndss-paper/peering-inside-the-black-box-long-range-and-scalable-model-architecture-snooping-via-gpu-electromagnetic-side-channel/}
}

@inproceedings{horvath2026kraken,
  title     = {Kraken: Higher-order {EM} Side-Channel Attacks on {DNNs} in Near and Far Field},
  author    = {Horv{\'a}th, P{\'e}ter and Shumailov, Ilia and Chmielewski, Lukasz and Batina, Lejla and Yarom, Yuval},
  booktitle = {IEEE Conference on Secure and Trustworthy Machine Learning (SaTML)},
  year      = {2026},
  eprint    = {2603.02891},
  archivePrefix = {arXiv},
  url       = {https://arxiv.org/abs/2603.02891}
}

@inproceedings{ding2025moecho,
  title     = {{MoEcho}: Exploiting Side-Channel Attacks to Compromise User Privacy in Mixture-of-Experts {LLMs}},
  author    = {Ding, Ruyi and Xu, Tianhong and Shen, Xinyi and Ding, Aidong Adam and Fei, Yunsi},
  booktitle = {Proceedings of the 2025 ACM SIGSAC Conference on Computer and Communications Security (CCS '25)},
  pages     = {2159--2173},
  year      = {2025},
  address   = {Taipei, Taiwan},
  publisher = {ACM},
  doi       = {10.1145/3719027.3765174}
}

@inproceedings{nuriyev2026expert,
  title     = {Expert Selections In {MoE} Models Reveal (Almost) As Much As Text},
  author    = {Nuriyev, Amir and Kulp, Gabriel},
  booktitle = {ICLR 2026 Workshop on Principled Design for Trustworthy AI},
  year      = {2026},
  eprint    = {2602.04105},
  archivePrefix = {arXiv},
  primaryClass  = {cs.CL},
  doi       = {10.48550/arXiv.2602.04105},
  url       = {https://arxiv.org/abs/2602.04105}
}

@inproceedings{sehatbakhsh2019emma,
  title     = {{EMMA}: Hardware/Software Attestation Framework for Embedded Systems Using Electromagnetic Signals},
  author    = {Sehatbakhsh, Nader and Nazari, Alireza and Khan, Haider and Zaji{\'c}, Alenka and Prvulovic, Milos},
  booktitle = {Proceedings of the 52nd Annual IEEE/ACM International Symposium on Microarchitecture (MICRO-52)},
  pages     = {983--995},
  year      = {2019},
  address   = {New York, NY, USA},
  publisher = {ACM}
}

@inproceedings{xiao2023magtracer,
  title     = {{MagTracer}: Detecting {GPU} Cryptojacking Attacks via Magnetic Leakage Signals},
  author    = {Xiao, Rui and Li, Tianyu and Ramesh, Soundarya and Han, Jun and Han, Jinsong},
  booktitle = {Proceedings of the 29th Annual International Conference on Mobile Computing and Networking (MobiCom '23)},
  pages     = {1--15},
  year      = {2023},
  publisher = {ACM},
  doi       = {10.1145/3570361.3613283}
}

@misc{sidechannelcloud,
  title        = {Side {C}hannel {C}loud: Instrumented {GPU} {I}nfrastructure for {AI} {S}ecurity {R}esearch},
  author       = {{Amodo Design}},
  howpublished = {\url{https://sidechannel.cloud}},
  year         = {2026}
}

@misc{dataset,
  author       = {Gargiulo, Simone},
  title        = {{NVIDIA} {H200} power side-channel traces of {GPU} training, inference and non-{AI} workloads},
  year         = {2026},
  howpublished = {Hugging Face dataset,
                  \url{https://huggingface.co/datasets/simgar/h200-power-workload-traces-10MHz_5s}},
  note         = {930 genuine and 680 adversarial five-second recordings over seventeen open models and twenty-five non-AI workloads. Nominal acquisition rate is \SI{10}{\mega\hertz}.}
}

@misc{micsig_rcp,
  author       = {{Shenzhen Micsig Technology}},
  title        = {Rogowski {AC} current probe, {RCP-XS} series},
  year         = {2026},
  howpublished = {Product page,
                  \url{https://en.micsig.com/product_detail/Flexible-Current-Probe--Rogowski-Coil-RCP-XS.html}},
  note         = {Accessed August 2026. The RCP120 variant used here reports 50~mV/A.}
}

@misc{DARPA_LADS,
  author       = {{DARPA-LADS}},
  title        = {Leveraging the {A}nalog {D}omain for {S}ecurity ({LADS})},
  year         = {2015},
  howpublished = {\url{https://www.darpa.mil/research/programs/leveraging-the-analog-domain-for-security}},
  note         = {Accessed: 2026-08-24}
}

\clearpage

\appendix

\section{Appendix}
\label{app:Appendix}

\begin{table}[!htpb]
\caption{\textbf{Throughput penalty.} We report for each evasion strategy the reduction in real training throughput relative to a full fine-tuning of the same model, defined as $\text{penalty} = 100\left(1-\frac{\text{evasion tokens/s}}{\text{full fine-tuning tokens/s}}\right)$. A positive penalty indicates slower training: for example, $+69.3\%$ indicates that the strategy achieves $30.7\%$ of the full fine-tuning training token rate. LoRA is generally much faster than full fine-tuning, and the penalty observed arises from the decoy inference interleaved with the training updates.}

\label{tab:penalty}
\centering
\begin{tabular}{lcccc}
\toprule
 & \multicolumn{4}{c}{Throughput penalty (\%)} \\
Model & Chunk. & Thrott. & Dilute & Dil. LoRA \\
\midrule
Qwen3 4B     & $+55.8$ & $+39.1$ & $+56.2$ & $+69.3$ \\
Llama 3.1 8B & $+54.7$ & $+39.5$ & $+54.9$ & $+64.7$ \\
Qwen3 14B    & $+34.1$ & $+39.4$ & $+54.9$ & $+61.9$ \\
GPT-OSS 20B  & $+28.1$ & $+41.7$ & $+48.1$ & $+33.8$ \\
\bottomrule
\end{tabular}
\end{table}

\begin{figure}[!htbp]
\includegraphics[width=\columnwidth]{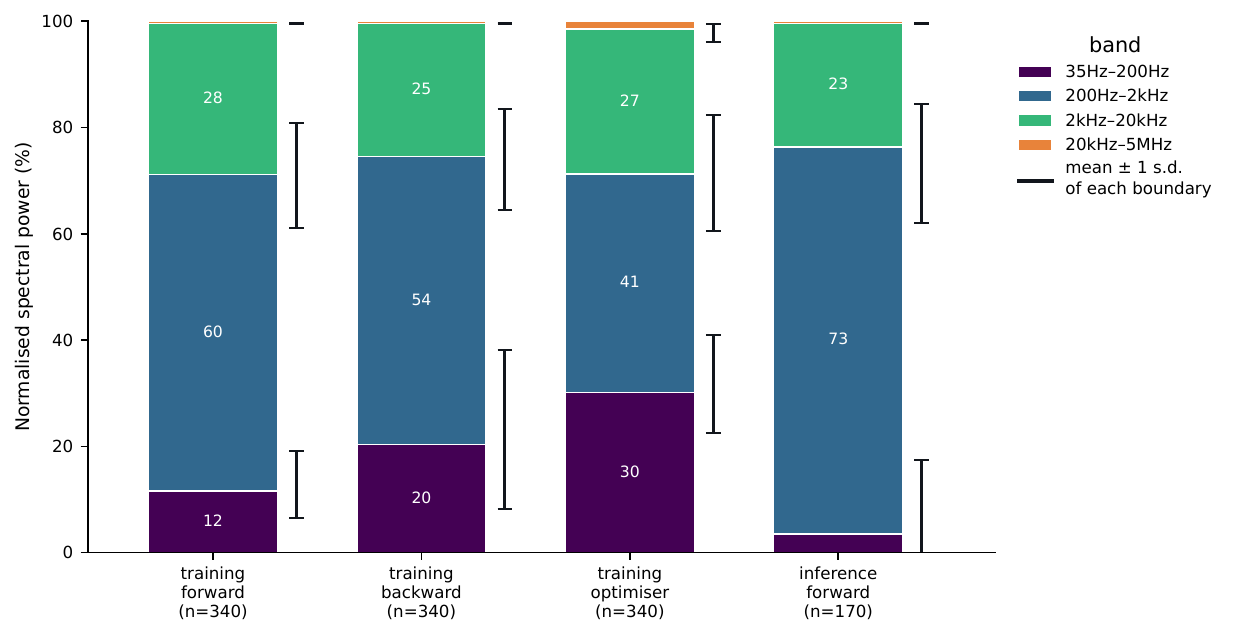}
\caption{Normalized spectral power per frequency band for three training phases (forward pass, backward pass and optimizer update) and for inference, expressed as a percentage of the total power within each phase. The probe is AC-coupled from \SI{34}{\hertz} to \SI{30}{\mega\hertz}, with a nominal sampling rate of \SI{10}{\mega\hertz}. The figure is computed from $510$ recordings of the genuine corpus: $340$ for training (including full fine-tuning and low-rank adaptation across all seventeen models) and $170$ for inference. Gradient accumulation has been excluded from this evaluation. Power is normalized in each recording so its frequency bands sum to \SI{100}{\percent}. Bars show the median share across all recordings and vertical marks indicate one standard deviation. The three frequency decades above \SI{20}{\kilo\hertz} have been merged for readability. Overall, they account for $0.41\%$, $0.45\%$, $1.47\%$, and $0.39\%$ of the total power in the four phases, respectively. Almost all spectral power lies below \SI{20}{\kilo\hertz}, while the phases separate most clearly below \SI{200}{\hertz}.}
\label{fig:bands}
\end{figure}

\begin{table}[!htbp]
\centering
\caption{Accuracy and macro-F1 as a function of sampling frequency, evaluated on the $930$ captures of the genuine corpus. Five-fold cross-validation grouped by model and shuffled, with class-balanced two-stage forests, averaged over four splits and two seeds. Performance is evaluated over the three classes (training, inference, non-AI) on individual \SI{5}{\second} traces. The majority-class baseline accuracy is $0.548$. At each rate, the bands above the corresponding Nyquist frequency are dropped, so the feature count falls from $30$ at the top rate to $24$ at the lowest.}
\label{tab:rates}
\begin{tabular}{lcc}
\toprule
Rate (MS/s) & {Accuracy} & {Macro-F1} \\
\midrule
 9.766 & {$0.966$} & {$0.955$} \\
 4.883 & {$0.970$} & {$0.961$} \\
 2.441 & {$0.966$} & {$0.955$} \\
 1.221 & {$0.958$} & {$0.944$} \\
 0.610 & {$0.968$} & {$0.959$} \\
 0.305 & {$0.961$} & {$0.948$} \\
 0.153 & {$0.949$} & {$0.930$} \\
 0.076 & {$0.921$} & {$0.901$} \\
 0.038 & {$0.946$} & {$0.928$} \\
\bottomrule
\end{tabular}
\end{table}

\begin{table}[!htbp]
\caption{\textbf{Features.} The two-stage random forest uses $30$ features, separated into the frequency and time domains. Features do not depend on absolute signal amplitudes. The envelope is the root-mean-square current over \SI{2}{\milli\second} frames. The frame is considered quiet when the envelope drops below $20\%$ of its floor to peak range ($1$st--$99$th percentiles), and active otherwise.}

\label{tab:features}
\centering
{\small
\begin{tabular}{llrl}
\toprule
Feature & Domain & Count & What it measures \\
\midrule
\texttt{band00}--\texttt{band13} & frequency & 14 & normalized spectral power in each log-spaced band, \\
                                 &           &    & \SI{20}{\hertz} to \SI{8}{\mega\hertz} \\
\texttt{sp\_centroid}            & frequency & 1  & spectral centroid on the log-frequency axis \\
\texttt{sp\_spread}              & frequency & 1  & spectral spread about the centroid \\
\texttt{sp\_skew}                & frequency & 1  & spectral skewness \\
\texttt{sp\_entropy}             & frequency & 1  & spectral entropy of the power distribution \\
\texttt{sp\_roll85}, \texttt{sp\_roll95} & frequency & 2 & frequencies containing $85\%$ and $95\%$ of total power \\
\texttt{frac\_below\_2k}         & frequency & 1  & fraction of total power below \SI{2}{\kilo\hertz} \\
\texttt{frac\_above\_200k}       & frequency & 1  & fraction of total power above \SI{200}{\kilo\hertz} \\
\midrule
\texttt{env\_cv}                 & time      & 1  & coefficient of variation of the envelope \\
\texttt{env\_crest}              & time      & 1  & envelope crest factor, peak-to-RMS ratio \\
\texttt{env\_rep\_hz}            & time      & 1  & repetition frequency estimated from envelope autocorrelation \\
\texttt{env\_rep\_strength}      & time      & 1  & autocorrelation strength at the repetition period \\
\texttt{env\_acf\_decay}         & time      & 1  & decay of the envelope autocorrelation \\
\texttt{env\_idle\_frac}         & time      & 1  & quiet fraction of the window \\
\texttt{env\_duty}               & time      & 1  & ratio of median active-run duration to median cycle duration \\
\texttt{env\_trans\_hz}          & time      & 1  & rate of active-to-quiet transitions \\
\midrule
\multicolumn{2}{l}{\emph{total}} & \textbf{30} & $22$ frequency-domain, $8$ time-domain \\
\bottomrule
\end{tabular}}
\end{table}

\begin{table}[!htbp]
\centering
\caption{\textbf{Composition of the dataset and list of the LLM models used for the experiment.} The genuine corpus contains $93$ labels and $930$ recordings, with $10$ repetitions per label. For each model, we record traces under inference and three training configurations: full fine-tuning, low-rank adaptation, and gradient accumulation, each contributing $170$ traces. Settings for the gradient-accumulation acquisitions are reported in Table~\ref{tab:accumsettings}. For the remaining workloads, each model uses the largest batch size possible so that a full fine-tuning step fits within the GPU memory: $8$ for thirteen models, $4$ for \texttt{gemma9b}, \texttt{nemo12b}, and \texttt{ossmoe20b}, and $2$ for \texttt{starcoder15b}. Sequence length is $512$ tokens. Inference workloads are prefill-style forward passes at the same batch size as training, with no decoding. Per model, inference and training recordings differ only in the task they perform.}

\label{tab:families}
{\small
\setlength{\tabcolsep}{4pt}
\begin{tabular}{llcccccc}
\toprule
Models & Identifier & Size & Arch. & Inference & Full FT & LoRA & Accum. \\
\midrule
qwen4b       & Qwen/Qwen3-4B                & 4B  & dense & $\bullet$ & $\bullet$ & $\bullet$ & $\bullet$ \\
mistral7b    & mistralai/Mistral-7B-v0.3    & 7B  & dense & $\bullet$ & $\bullet$ & $\bullet$ & $\bullet$ \\
qwen8b       & Qwen/Qwen3-8B                & 8B  & dense & $\bullet$ & $\bullet$ & $\bullet$ & $\bullet$ \\
apertus8b    & swiss-ai/Apertus-8B-2509     & 8B  & dense & $\bullet$ & $\bullet$ & $\bullet$ & $\bullet$ \\
granite8b    & ibm-granite/granite-3.3-8b-base   & 8B  & dense & $\bullet$ & $\bullet$ & $\bullet$ & $\bullet$ \\
llama8b      & meta-llama/Llama-3.1-8B      & 8B  & dense & $\bullet$ & $\bullet$ & $\bullet$ & $\bullet$ \\
yi9b         & 01-ai/Yi-1.5-9B              & 9B  & dense & $\bullet$ & $\bullet$ & $\bullet$ & $\bullet$ \\
gemma9b      & google/gemma-2-9b            & 9B  & dense & $\bullet$ & $\bullet$ & $\bullet$ & $\bullet$ \\
falcon10b    & tiiuae/Falcon3-10B-Base      & 10B & dense & $\bullet$ & $\bullet$ & $\bullet$ & $\bullet$ \\
nemo12b      & mistralai/Mistral-Nemo-Base-2407  & 12B & dense & $\bullet$ & $\bullet$ & $\bullet$ & $\bullet$ \\
stablelm12b  & stabilityai/stablelm-2-12b   & 12B & dense & $\bullet$ & $\bullet$ & $\bullet$ & $\bullet$ \\
olmo13b      & allenai/OLMo-2-1124-13B      & 13B & dense & $\bullet$ & $\bullet$ & $\bullet$ & $\bullet$ \\
qwen14b      & Qwen/Qwen3-14B               & 14B & dense & $\bullet$ & $\bullet$ & $\bullet$ & $\bullet$ \\
phi4         & microsoft/phi-4              & 14B & dense & $\bullet$ & $\bullet$ & $\bullet$ & $\bullet$ \\
starcoder15b & bigcode/starcoder2-15b       & 15B & dense & $\bullet$ & $\bullet$ & $\bullet$ & $\bullet$ \\
dsv2lite     & deepseek-ai/DeepSeek-V2-Lite & 16B & MoE   & $\bullet$ & $\bullet$ & $\bullet$ & $\bullet$ \\
ossmoe20b    & openai/gpt-oss-20b           & 21B & MoE   & $\bullet$ & $\bullet$ & $\bullet$ & $\bullet$ \\
\midrule
\multicolumn{8}{l}{Non-AI ($25$ labels): idle, matmul\_fp16, matmul\_bf16, matmul\_tall, matmul\_int8, cufft\_c2c, cufft\_r2c,} \\
\multicolumn{8}{l}{\quad   conv2d, reduce\_sum, sort, cumsum, transpose, elementwise, memcpy\_d2d, gather\_random, } \\
\multicolumn{8}{l}{\quad scatter\_random, rng\_normal, histogram, topk, softmax\_large, layernorm\_only, l2\_thrash, } \\
\multicolumn{8}{l}{\quad  atomics\_contended, spmv\_csr, pcie\_h2d} \\
\bottomrule
\end{tabular}}
\end{table}

\begin{table}[!htbp]
\centering
\caption{\textbf{Workload types}. Genuine workloads are separated into three classes: inference, training, and non-AI activities. The adversarial corpus is separated into four categories named after the manipulation they apply to the training schedule. For example, chunked optimization splits the optimizer update into smaller portions interleaved with inference passes, while in diluted LoRA the optimizer update occupies only a small fraction of the acquisition window. Every workload runs in loop across the recording, so the active fraction is $1.0$ for every class.}
\label{tab:workloads}
\begin{tabular}{llcc}
\toprule
Type & Class & Models & Recordings \\
\midrule
inference             & genuine   & 17 & 170 \\
full fine-tuning      & genuine   & 17 & 170 \\
low-rank adaptation (LoRA)   & genuine   & 17 & 170 \\
gradient accumulation & genuine   & 17 & 170 \\
non-AI kernels        & genuine   & 25 labels & 250 \\
\midrule
chunked optimizer     & adversarial & {\boldmath17} & 170 \\
throttling            & adversarial & {\boldmath17} & 170 \\
dilution              & adversarial & {\boldmath17} & 170 \\
diluted LoRA          & adversarial & {\boldmath17} & 170 \\
\bottomrule
\end{tabular}
\end{table}

\begin{table}[!htbp]
\caption{\textbf{Gradient accumulation settings.} In gradient accumulation, several micro-batches are processed before a single optimizer update. By default, each model accumulates for $4$ micro-batches, with a batch size specified in Table~\ref{tab:families}. Here, we list the three models that need to depart from this setting. For these models, the number of micro-batches needs to be reduced so that one complete accumulation cycle fits within the \SI{5}{\second} recording window: \texttt{ossmoe20b} only fits three, while \texttt{olmo13b} and \texttt{qwen14b} fit two. In addition, batch size is reduced independently when the accumulated cycle would otherwise exceed device memory. Step time is the duration of one complete accumulation cycle, including the optimizer update.}
\label{tab:accumsettings}
\centering
\begin{tabular}{llcccc}
\toprule
Model & Size & Architecture & Step (s) & Micro-batches & Batch \\
\midrule
\texttt{ossmoe20b} & 21B & MoE ($4/32$ active) & $1.06$ & $3$ & $2$ \\
\texttt{olmo13b}   & 13B & dense               & $1.24$ & $2$ & $4$ \\
\texttt{qwen14b}   & 14B & dense               & $1.24$ & $2$ & $8$ \\
\bottomrule
\end{tabular}
\end{table}

\subsection{Diluted LoRA rescue rule}
\label{app:rescue}

Diluted LoRA hides an optimizer update within a long stream of decoy inference, so the update occupies only about $1\%$ of the acquisition window. Since the two-stage random forest computes its features over the entire recording, the short-lived training fingerprint is not able to substantially shift their values, reducing the performance of the detector. 

The training signature, however, remains quite distinct in the time-frequency domain, wherever the optimizer update occurs (for the configurations under study). It is quieter than the surrounding phases and its spectrum is weighted towards lower frequencies. The rescue rule targets this event directly.

We compute the envelope $e$ as the root-mean-square current over \SI{0.5}{\milli\second} frames and define the dip depth as

\begin{equation}    
  \texttt{deep\_dip}
  = 1 - \frac{\min_t \bar{e}_{20}(t)}
  {\operatorname{median}(e)},
\end{equation}

where $\bar{e}_{20}(t)$ is the $\SI{20}{\milli\second}$ rolling mean of $e$. Over a \SI{5}{\second} recording, both $e$ and $\bar{e}_{20}(t)$ form time series of approximately \num[print-unity-mantissa=false]{1e4} values. Forward and backward passes are relatively flat, giving $\texttt{deep\_dip}\sim 0$, while the optimizer update generally produces a larger value. In this way, a single update event can remain observable even when surrounded by inference.

Depth threshold, however, may also be activated by an idle or throttled GPU event. Therefore, we introduce a frequency gate on the spectral content of the dip. The quantity $\texttt{lowfreq}$ is computed over the \SI{20}{\milli\second} window centered at its deepest point and represents the normalized fraction of spectral power below \SI{10}{\kilo\hertz}. The condition $\texttt{lowfreq}>0.5$ separates dips produced by the optimizer update from idle or quiescent phases. The optimizer update has most of its spectral share at low frequencies, whereas idles are dominated by high-frequency noise and a tone at \SI{500}{\kilo\hertz}.
Finer spectral gating and rescue rules could be defined using the information in Fig.~\ref{fig:bands}.

When the second stage has labelled a recording as inference, the rescue rule can change the labelling and flag it as training if

\begin{equation}
\texttt{deep\_dip} > \tau
\quad\text{and}\quad
\texttt{lowfreq} > 0.5.
\end{equation}

The threshold $\tau$ is calibrated using genuine recordings alone. Candidate values are taken from the \texttt{deep\_dip} scores of genuine training traces and tested on inference and non-AI recordings. The threshold is set to the smallest value that does not increase the detector's false-alarm rate on these recordings. In other words, $\tau$ is chosen as the smallest threshold for which none of the inference and non-AI recordings of the genuine corpus are relabelled as training by the rescue rule. Therefore, by construction, the rule does not add false alarms on the genuine corpus. The threshold is computed separately for each setting of the classifier: $0.40$, $0.32$, and $0.47$ for the genuine, unseen-strategy, and unseen-model detectors, respectively. These values decrease to $0.38$, $0.31$ and $0.43$ when calibrated out-of-fold, respectively. Diluted LoRA traces typically reach \texttt{deep\_dip} values with a median of approximately $0.68$.

Applied only to the diluted LoRA arm, the rule improves detection, as reported in Table~\ref{tab:detectors}, with no additional in-sample false-alarm cost. This holds true by construction, since $\tau$ is calibrated on the same recordings on which the false alarm is then measured. To test whether the rule generalises, we recompute $\tau$ out-of-fold: it is calibrated on part of the genuine recordings, while the false alarm rate is evaluated on a held-out remainder. The split is performed by label for the genuine and unseen-strategy detectors, and by model for the unseen-model detector. Out-of-fold, the rescue rule adds $+0.6$, $+2.5$ and $+1.4$ percentage points to the false-alarm rate for the genuine, unseen-strategy and unseen-model detectors, respectively.

\end{document}